\documentclass[prl,floatfix,showpacs,twocolumn,amsmath,amssymb]{revtex4}
\usepackage{natbib}
\usepackage{dcolumn}
\usepackage{graphicx}
\usepackage{color}
\usepackage{ulem}

\newcommand{\be}{\begin{equation}}
\newcommand{\ee}{\end{equation}}
\newcommand{\bea}{\begin{eqnarray}}
\newcommand{\eea}{\end{eqnarray}}

\def\[{\begin{equation}}
\def\]{\end{equation}}
\begin{document}
\title{Stringent Restriction from the Growth of Large-Scale Structure on Apparent Acceleration in Inhomogeneous Cosmological Models}
\author{Mustapha Ishak\footnote{Electronic address: mishak@utdallas.edu}}
\author{Austin Peel}
\author{M. A. Troxel}
\affiliation{
Department of Physics, The University of Texas at Dallas, Richardson, TX 75080, USA}
\date{\today}
\begin{abstract}
Probes of cosmic expansion constitute the main basis for arguments to support or refute a possible apparent acceleration due to different expansion rates in the universe as described by inhomogeneous cosmological models.  We present in this \textit{Letter} a separate argument based on results from an analysis of the growth rate of large-scale structure in the universe as modeled by the inhomogeneous cosmological models of Szekeres. We use the models with no assumptions of spherical or axial symmetries. We find that while the Szekeres models can fit very well the observed expansion history without a $\Lambda$, they fail to produce the observed late-time suppression in the growth unless $\Lambda$ is added to the dynamics. A simultaneous fit to the supernova and growth factor data shows that the cold dark matter model with a cosmological constant ($\Lambda$CDM) provides consistency with the data at a confidence level of $99.65\%$ while the Szekeres model without $\Lambda$ achieves only a $60.46\%$ level. When the data sets are considered separately, the Szekeres with no $\Lambda$ fits the supernova data as well as the $\Lambda$CDM does, but provides a very poor fit to the growth data with only $31.31\%$ consistency level compared to $99.99\%$ for the $\Lambda$CDM. This absence of late-time growth suppression in inhomogeneous models \textit{without} a $\Lambda$ is consolidated by a physical explanation.        
\end{abstract} 
\maketitle
%
%
%
\setlength{\belowdisplayskip}{1.2pt} \setlength{\belowdisplayshortskip}{1.2pt}
\setlength{\abovedisplayskip}{1.2pt} \setlength{\abovedisplayshortskip}{1.2pt}
{\parindent0pt\it \textbf{Introduction.}}
%
We are witnessing a flourishing era in cosmology where complementary observations and data sets are becoming available at an impressive rate. During the last few decades, we have learned and confirmed a great deal of knowledge about our universe from analyzing these data sets, including its age, dynamics, and evolution. However, this progress has also come with two outstanding conundrums. One is the problem of the dark matter, which manifests itself via its gravitational pull, while the second is the cosmic acceleration or dark energy problem, which indicates repulsive gravitational dynamics at large distance scales in the universe. We are interested in the latter problem here. 

At least three possible causes have been proposed by the scientific community in order to try to explain the source of cosmic acceleration. The first is the presence of a prevalent cosmological constant or dark energy component permeating the universe. The second possibility is that cosmic acceleration is due to an extension or modification to general relativity that takes effect at cosmological scales. A third possibility put forward in the scientific literature is that we live in a lumpy universe, where observations are affected by inhomogeneities and require more elaborate functions to describe them using relativistic inhomogeneous models. Such altered observations can lead to an apparent acceleration due to different  Hubble expansions from one region of the universe to another. For example, an underdense region will be subject to less gravitational pull from its matter content, and thus it expands faster than the global average. An observer located in such a region will observe the surrounding universe outside that region to recede faster than his local region. Such an effect cannot be captured when the Hubble function is only a function of time, as in the Friedmann-Lema\^itre-Robertson-Walker (FLRW) models, where the only possible interpretation then becomes a true acceleration. In inhomogeneous cosmological models, the Hubble function depends on space and time, so a spatial variation is allowed without necessarily inferring any acceleration. This is the main idea of apparent acceleration, and there have been pros and cons for it. See \cite{Ellis1} for a review.

Previous analyses with inhomogeneous models have mainly focused on probes of the expansion history, such as supernova luminosity distance-redshift relations, angular distance to the cosmic microwave background last scattering surface, and angular diameter distance and Hubble expansion in baryon acoustic oscillations. Also, these studies mostly used the spherically symmetric Lema\^itre-Tolman-Bondi (LTB) models, where observations restrict the observer to be close to its center and thus violate the Copernican principle. For example, see \cite{Ellis1} and references therein. The kinetic Sunayev-Zeldovich effect and CMB full analysis have also put some challenging constraints on the LTB models, e.g. \cite{ZhangStebbins,MossZibinScott}.
 
In this \textit{Letter}, we address the question of apparent acceleration using an argument based on the growth rate of large-scale structure (the formation history of clusters and superclusters of galaxies) in the universe. We use the Szekeres inhomogeneous cosmological models, which have no artificial symmetry \cite{Szekeres1,Szekeres2} and are compatible with the Copernican principle \cite{Ishaketal2008}. These models are exact solutions to Einstein's equations solved with no symmetries for an irrotational dust source. They can represent a lumpy universe filled with underdense and overdense regions, and they are regarded as the best known exact solutions one can use for these types of studies \cite{Ellis2}. 
\\
\indent It is worth noting that independent of distances to supernovae, strong support for cosmic acceleration and the need for a cosmological constant came from observations of galaxy cluster properties as functions of redshift. These have been joined by CMB or baryon acoustic oscillation measurements that also independently support an acceleration. Besides the usual cluster number counts indicating a low density universe, there is also an argument based on the time evolution of the size and abundance of clusters,  which is of interest to us here. Indeed, it was shown, e.g. \cite{Allen,Vik}, that the growth rate of formation of clusters is suppressed at late times, and that a cosmological constant is necessary in order to explain this late-time suppression in FLRW models. Similarly, other probes of the large-scale growth factor using redshift space distortions and Lyman-$\alpha$ forest also show the presence of this late-time suppression \cite{Guzzo2008,Blake,Viel,Boss1}, thus concurring with cluster observations. As we will show, the Szekeres inhomogeneous models that are able to fit the observed expansion history do not exhibit the necessary late-time suppression of growth if we do not include a cosmological constant.
\\[2pt]
{\parindent0pt\it \textbf{Fitting the expansion history to the Szekeres models.}}
The study of the expansion history and cosmological distances in the Szekeres models can be best visualized by using their LTB-like representation, since it allows for an easier comparison of their geometry to that of the well-known LTB and FLRW models. The metric is \cite{HK}
\be
ds^2= -\mathrm{d}t^2+\frac{(\Phi_{,r}-\Phi E_{,r}/E)^2}{\epsilon-k}\mathrm{d}r^2 +\frac{\Phi^2}{E^2}(\mathrm{d}p^2+\mathrm{d}q^2),
\label{eq:metric}
\ee
where a comma denotes partial differentiation, $\Phi=\Phi(t,r)$ is analog to an areal radius, $k=k(r)$ determines the curvature of the $t=$ const. spatial sections, and $E=E(r,p,q)$ determines the mapping of the coordinates $(p, q)$ onto the 2-space for each value of $r$. It is given by 
\be
E(r,p,q)=\frac{S(r)}{2}[(\frac{p-P(r)}{S(r)})^2+(\frac{q-Q(r)}{S(r)})^2 + \epsilon],
\ee
where $S$, $P$, and $Q$ are arbitrary functions of $r$. The constant $\epsilon$ determines whether the $(p,q)$ 2-surfaces are 
spherical ($\epsilon=+1$), pseudo-spherical ($\epsilon=-1$), or planar ($\epsilon=0$)---that is, it controls how the 2-surfaces of constant $r$ foliate the 3-dimensional spatial sections of constant $t$. The Einstein field equations with a $\Lambda$ give
\be
(\Phi_{,t})^2={2M}/{\Phi}-k
\label{eq:EE1}
\ee
\be
8 \pi \rho(t,r,p,q)=\frac{2(M_{,r}-3 M E_{,r}/E)}{\Phi^2(\Phi_{,r}-\Phi E_{,r}/E)},
\label{eq:EE2}
\ee
where $M(r)$ represents the total active gravitational mass in the case $\epsilon =+1$ \cite{HK}, and we use units where $c=G=1$. The evolution of $\Phi(t,r)$ divides the models into 3 sub-cases: hyperbolic $(k(r)<0)$, parabolic $(k(r)=0)$, and elliptic $(k(r)>0)$.
The spherically symmetric (LTB) subcase results when {$E_{,r}=0$}, and the dust FLRW arises when $\Phi(t,r)=a(t)r$ and $k(r)=k_0 r^2$, where $a(t)$ is the scale factor, and $k_0$ the spatial curvature index.

Now, in order to calculate distances and redshift in a cosmological model, one has to solve the null geodesic equations that govern the propagation of light rays. In inhomogeneous models, one has to use numerical integrations, since these equations are not integrable analytically. Additionally, one has to solve for the optical scalar equations \cite{Sachs} or for the partial derivatives of the null vector components \cite{Sachs,NIT2011}. We use the affine null geodesic equations ${k}^{\alpha}{}_{;\beta} {k}^{\beta}=0$ that read 
\be
\dot k^t + \frac{\mathcal{H}_{,t}}{2}(k^r)^2+\frac{{\mathcal{F}^2}_{ ,t}}{2}[(k^p)^2+(k^q)^2]=0 \label{geodkt2},
\ee
\be
\mathcal{H}\dot k^r+\dot{\mathcal{H}}k^r-\frac{\mathcal{H}_{,r}}{2}(k^r)^2-\frac{{\mathcal{F}^2}_{ ,r}}{2}\big{[}(k^p)^2+(k^q)^2\big{]}=0\label{geodkr2},
\ee
\be
\mathcal{F}^2\dot k^p -\frac{\mathcal{H}_{,p}}{2}(k^r)^2+(\mathcal{F}^2){\bf \dot{}}\,k^p-\frac{{\mathcal{F}^2}_{ ,p}}{2}\big{[}(k^p)^2+(k^q)^2\big{]}=0,\label{geodkp2}
\ee
\be
\mathcal{F}^2\dot k^q -\frac{\mathcal{H}_{,q}}{2}(k^r)^2+(\mathcal{F}^2){\bf \dot{}}\,k^q-\frac{{\mathcal{F}^2}_{,q}}{2}\big{[}(k^p)^2+(k^q)^2\big{]}=0,\label{geodkq2}
\ee
\begin{figure*}
\begin{center}
\begin{tabular}{c c c}
{\includegraphics[width=5.7cm,height=5.2cm]{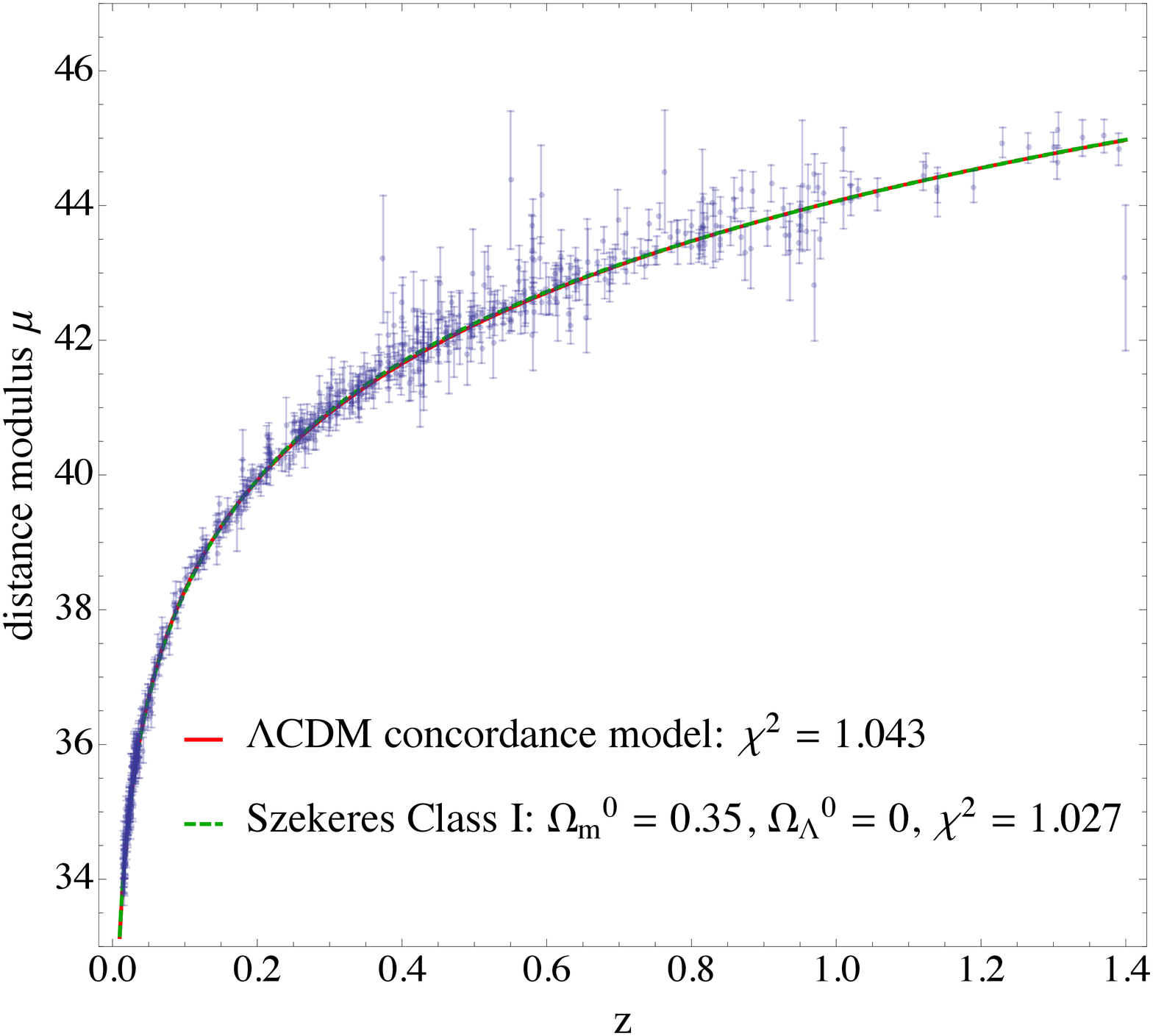}}&
{\includegraphics[width=5.7cm,height=5.2cm]{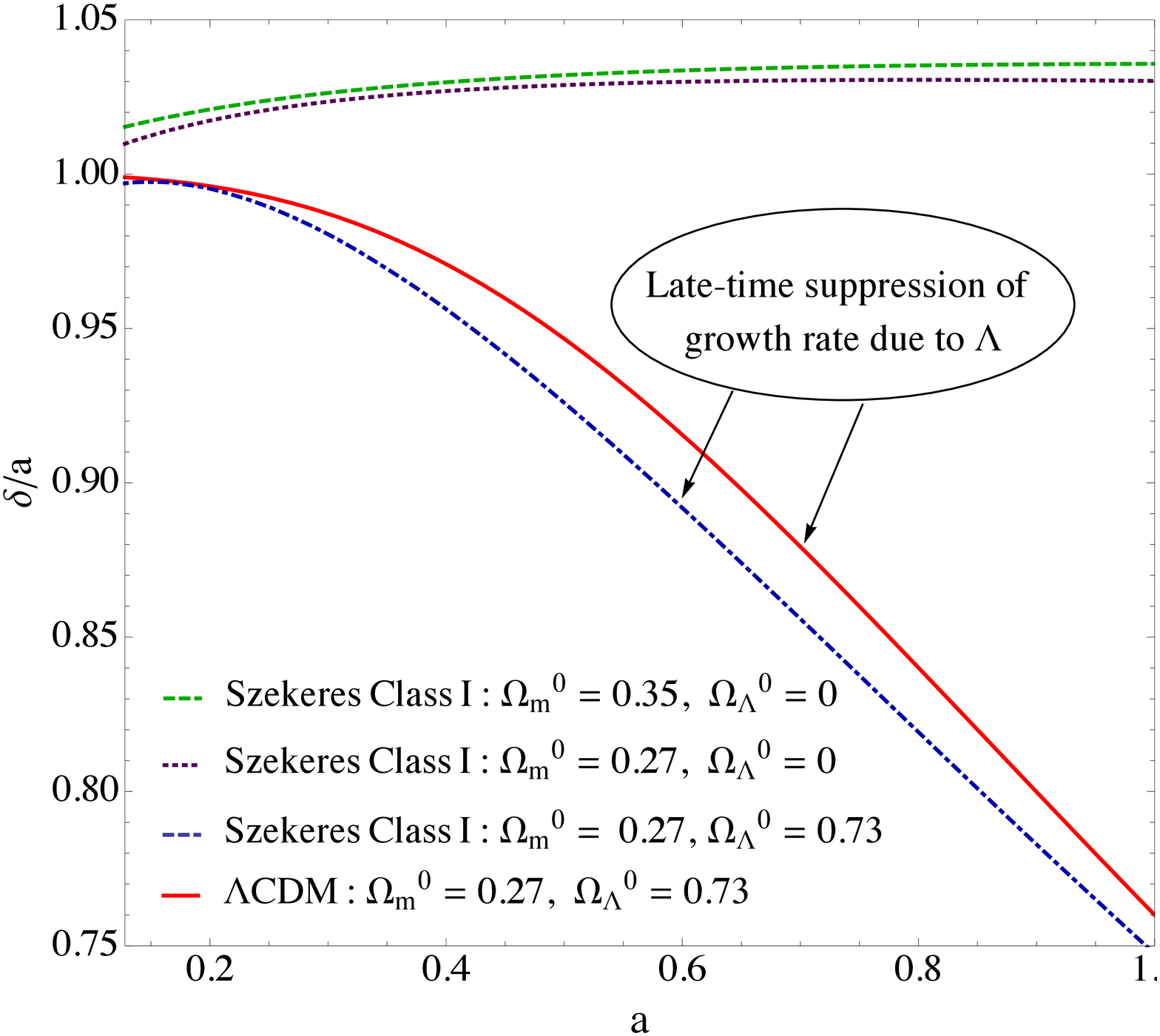}}&
{\includegraphics[width=5.7cm,height=5.2cm]{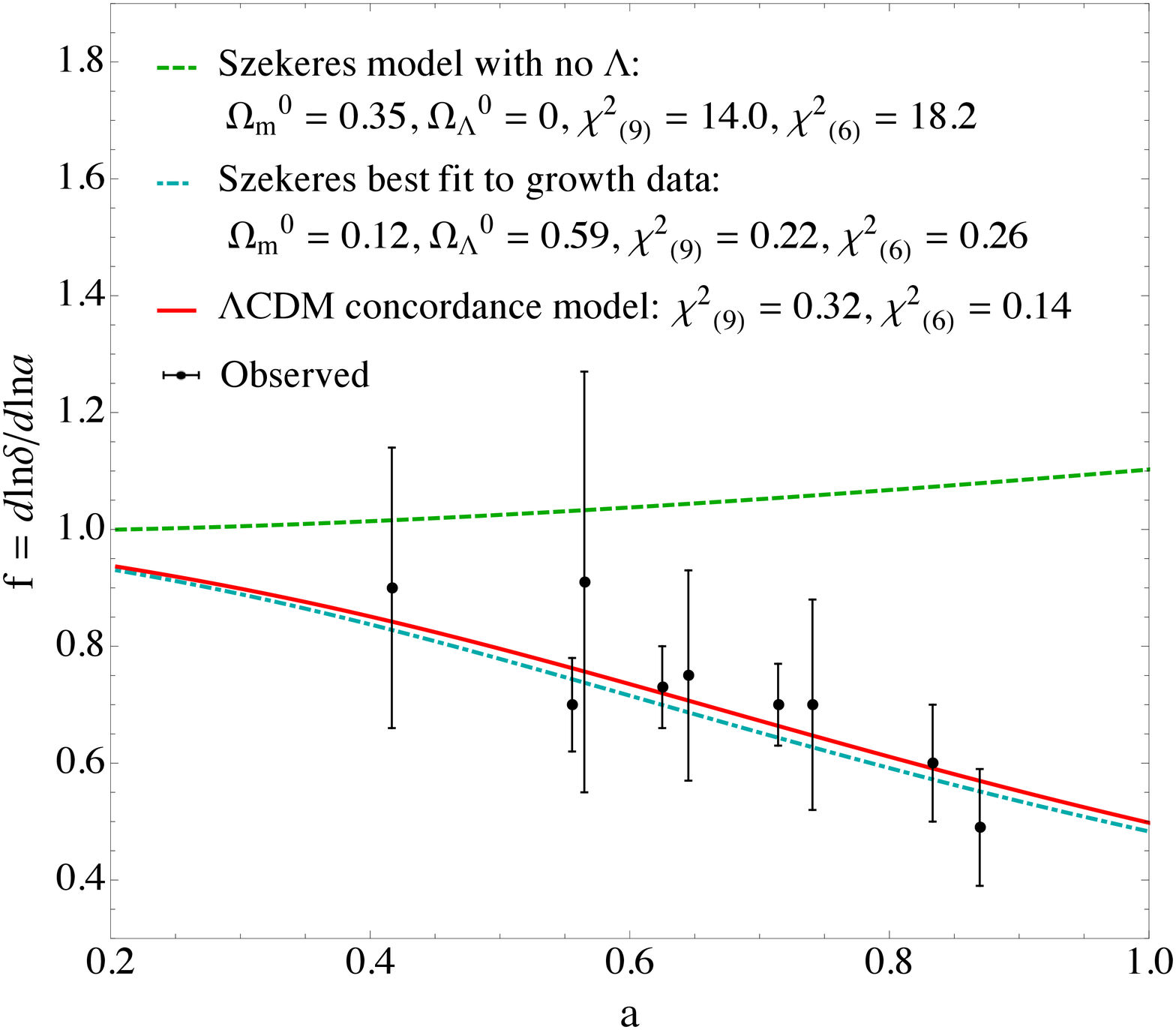}} \\
\end{tabular}
\setlength{\abovecaptionskip}{-0.1cm}
\setlength{\abovecaptionskip}{-0.1cm}
\setlength{\belowcaptionskip}{-1.cm}
\caption{\label{fig:luminosity}
LEFT: Distance modulus-redshift plots for the Szekeres Class I model with no cosmological constant, $\Lambda$, and the FLRW-$\Lambda$CDM concordance model, along with supernova Union2 compilation data \cite{Aman}. The very close normalized $\chi^2$ values and the small difference in magnitude between the two models (i.e. $0.00 \le |\mu_{_{\Lambda CDM}}- \mu_{_{Szekeres}}| \le 0.073$) show that practically the two models fit equally well the supernova data. CENTER: Plots of the growth rate of large-scale structure for Szekeres models with and without $\Lambda$, and the $\Lambda$CDM. The Szekeres model with a $\Lambda$ exhibits a late-time suppression of the growth similar to that of $\Lambda$CDM. Despite a very competitive fit to the supernova data, the Szekeres model with no $\Lambda$ exhibits no such late-time suppression. RIGHT: Plots of the growth factor and redshift space distortions data (e.g. \cite{Blake,Guzzo2008}). The Szekeres with $\Lambda$ and the $\Lambda$CDM are consistent with the growth data at $99.96\%$ and $99.99\%$ level of confidence, respectively. A simultaneous fit to the supernova and growth data gives $99.65\%$ and $60.46\%$ consistency for the $\Lambda$CDM and the Szekeres without $\Lambda$, respectively. The Szekeres with no $\Lambda$ that is best fit to the growth data provides only $31.31\%$ consistency level with the growth data.}
\end{center}
\end{figure*}
where $s$ is an affine parameter, ${k}^{\alpha}=dx^{\alpha}/ds\,\,$ is the null tangent vector, and $\dot {}=d/ds$. We have defined $\mathcal{F}\equiv{\Phi}/{E}$ and $\mathcal{H}\equiv{(\Phi_{,r}-\Phi E_{,r}/E)^2}/{(\epsilon-k)}$ for simplification. As done in our previous work, \cite{NIT2011}, we also use the 16 equations resulting from the partial derivatives of the 4 null geodesic equations to calculate the derivatives $k^\alpha{}_{,\beta}$.

We recall that the area distance $D_A$ relates to the surface area $\delta S$ of a propagating light front of a bundle of light rays by  $\delta S= D_A{}^2\delta \Omega$, where $\delta \Omega$ is the solid angle element. The surface area is also related to the expansion optical scalar $\Theta=\frac{1}{2} {k}^\alpha {}_{;\alpha}$ by the relation $d \ln(\delta S)=2 \Theta ds$. Combining these two equations gives \cite{Sachs}
\noindent
\bea
d\ln{D_A}&=&\Theta\,\,ds=({1}/{2})\,\,\, {k}^\alpha\;_{;\alpha}\,\,ds.
\label{eq:AreaDistance}
\eea
\noindent
After some steps, we obtain from Eqs. (\ref{eq:metric}) and (\ref{eq:AreaDistance}) the following expression for the area distance in Szekeres models 
\bea
d\ln{D_{A}}&=&\Big{[}\frac{1}{2} (\frac{{\mathcal{F}^2}_{,t}}{\mathcal{F}^2} k^t+\frac{{\mathcal{F}^2}_{,r}}{\mathcal{F}^2}  k^r+\frac{{\mathcal{F}^2}_{,p}}{\mathcal{F}^2}  k^p+\frac{{\mathcal{F}^2}_{,q}}{\mathcal{F}^2}  k^q )\nonumber\\
&+&\frac{1}{4} (\frac{\mathcal{H}_{,t}}{\mathcal{H}}  k^t+\frac{\mathcal{H}_{,r}}{\mathcal{H}}  k^r+\frac{\mathcal{H}_{,p}}{\mathcal{H}}  k^p+\frac{\mathcal{H}_{,q}}{\mathcal{H}}  k^q )\nonumber\\
&+&({1}/{2}) (k^t{}_{,t}+  k^r{}_{,r}+  k^p{}_{,p}+  k^q{}_{,q} ) \Big{]}ds.
\label{eq:diffareadistance}
\eea
The luminosity distance follows as $D_L=(1+z)^2 \,D_A$, and we use 
\be 
{\mu=m(z)-M=5 \log (D_L/\text{Mpc}) + 25}
\label{modulus}
\ee
for the distance modulus to a {supernova} of apparent and absolute magnitudes $m(z)$ and $M$, respectively. 
The redshift derives from the standard  relation \cite{Ellis2}
\begin{equation}
1+z={(  k^\alpha u_{\alpha})_e}/{(  k^\alpha u_{\alpha})_o}={(k^t)_e}/{(  k^t)_o},
\label{RedshiftStandard}
\end{equation}
where $u^{\alpha}=(1,0,0,0)$ is the 4-velocity vector of the comoving fluid, and the subscripts $e$ and $o$ stand for emitted and observed. 
Once explicit functions for a Szekeres model are specified, we numerically integrate the null geodesic equations (\ref{geodkt2})--(\ref{geodkq2}) along with their partial derivative equations in order to calculate the components of the null vectors, their partial derivatives, and the area distance. The distance modulus and the redshift then follow from Eqs.(\ref{modulus}) and (\ref{RedshiftStandard}) to plot Hubble diagrams for the expansion history in the Szekeres models. 

We use model functions that are similar to those used in our previous work \cite{Bolejko2006,PIT2012}, resulting in an inhomogeneous region approximately one hundred Mpc across that is representative of a supercluster of galaxies and a void surrounded by an almost FLRW model:
$\epsilon=+1$,
$M(r)=(4\pi/3)\rho_b\,r^3(1-\exp[-3(r/\sigma)^3])$, 
and $\{S,P,Q\}=\{140,\,10,\, -113\,\ln(1+r)\}$,
where $\sigma=30$, $\rho_b=(3/8\pi)H_0{}^2\Omega_m^0$, and $k(r)$ is calculated by integrating Eq.(\ref{eq:EE1}) with the coordinate choice $\Phi=r$ today. Another arbitrary function, $t_B(r)$, called the bang time arises from integrating the generalized Friedmann equation and is set here to be a constant in the almost FLRW region. We use $H_0=72$ km s${}^{-1}$ Mpc${}^{-1}$. The distance modulus-redshift results are plotted in Fig.\ref{fig:luminosity}(a) against the Union2 SN Ia compilation data \cite{Aman}. The Szekeres model has a normalized $\chi^2=1.027$ and the $\Lambda$CDM model has $\chi^2=1.043$,  showing that the expansions of the two models are equally competitive. One can see from (\ref{eq:diffareadistance}) and (\ref{RedshiftStandard}) that the distance and redshift have an angular dependence in general for Szekeres models. The model used here, however, does not exhibit enough anisotropy to affect the distance modulus-redshift relation noticeably.
\\[5pt]
{\parindent0pt\it \textbf{The absence of late-time suppression of the growth rate of large-scale structure in the Szekeres inhomogeneous cosmological models.}}
For the study of the growth rate, the GW formulation of the models is more convenient \cite{GW1,GW2,PIT2012}. We briefly present it here and refer the reader to \cite{GW1,GW2}. The metric is 
\be
	ds^2= -\mathrm{d}t^2+a^2\left[e^{2\nu}(\mathrm{d}\tilde{x}^2+\mathrm{d}\tilde{y}^2)+H^2W^2 \mathrm{d}r^2\right],
\label{eq:metric2}
\ee
and the space dependence of the functions $a(t,r)$, $\nu(r,\tilde{x},\tilde{y})$, $H(t,r,\tilde{x},\tilde{y})$, and $W(r)$ define two classes for the models; $\tilde{x}$ and $\tilde{y}$ are coordinates resulting from stereographic projections \cite{Szekeres1,Szekeres2}. The specific forms of these dependencies can be found in \cite{GW1,GW2}, and their illustration is not necessary for our purpose. We will use here Class I, which is more general. The models' time-evolution is given by the generalized Friedmann equation
\be
	\frac{\dot{a}(t,r)^2}{a(t,r)^2}=\frac{2\tilde{M}(r)}{a(t,r)^3}-\frac{\tilde{k}}{a(t,r)^2}+\frac{\Lambda}{3},
\label{Friedmann}
\ee
and the following evolution equation 
\be
	\ddot{F}(t,r)+ 2\frac{\dot{a}(t,r)}{a(t,r)}\dot{F}(t,r)-\frac{3\tilde{M}(r)}{a(t,r)^3}F(t,r)=0,
\label{Ray1}
\ee
where now $\dot{}\equiv\partial/\partial t$. $a(t,r)$ and $\tilde{M}(r)$ play the role of a scale factor and an effective gravitational mass, respectively, and $F$ arises from the splitting $H=A-F$, with the function $A(r,\tilde{x},\tilde{y})$ specified according to class \cite{GW1,GW2}. 
The matter density is given by 
\be
	8\pi\rho(t,r,\tilde{x},\tilde{y})=\frac{6\tilde{M}A}{a^3H}=\frac{6\tilde{M}}{a^3}(1+\frac{F}{H} ).
\label{density}
\ee
The evolution equation (\ref{Ray1}) follows from the field equations, or alternatively from the Raychaudhuri evolution equation \cite{Raychaudhuri,Ellis2} for an irrotational dust with a cosmological constant that is given by  
\be
\dot{\Theta}+\Theta^2/3=-2 \sigma^2-\rho/2+\Lambda.
\label{Ray2}
\ee 
Here $\Theta$ is the expansion rate scalar, $\sigma$ is the shear rate scalar, and $\rho$ is the matter density. This equation represent gravitational attraction and clustering \cite{Ellis1}. As discussed in \cite{PIT2012} and initiated by \cite{GW1,GW2}, one can identify an exact density fluctuation $\hat\delta=F/H$, which measures exact deviations from some background density $\rho(t,r)={6\tilde{M}(r)}/{a^3(t,r)}$ to write Eq.(\ref{density}) as 
\be
	\rho(t,r,\tilde{x},\tilde{y})= \rho(t,r)[1+\hat\delta(t,r,\tilde{x},\tilde{y})].
\ee 
With these definitions, the time evolution Eq.(\ref{Ray1}) can be written as a meaningful growth evolution equation 
\be
	\ddot{\hat\delta}+2\frac{\dot{a}(t,r)}{a(t,r)}\dot{\hat\delta}-\frac{3\tilde{M}(r)}{{a^3(t,r)}}\hat\delta-\frac{2}{1+\hat\delta}{{\dot{\hat\delta}^2}}-\frac{3\tilde{M}(r)}{a^3(t,r)}\hat\delta^2=0.
\label{growth1_classI}
\ee
Now, following standard steps used for FLRW and LTB models, we write this equation in terms of cosmological parameters as evaluated today. Each surface of constant $t$ and $r$ in the models evolves independently, so we can fix $r$ at some $r_s$ and use $a(t,r_s)$ as the time parameter for that surface. Then, after some steps and using Eq.(\ref{Friedmann}), we can rewrite the growth equation Eq.(\ref{growth1_classI}) as 
\be
	\hat\delta''+(\frac{4+2\,\Omega_\Lambda-\Omega_m}{2a})\hat\delta'-\frac{3}{2}\frac{\Omega_m}{a^2}\hat\delta-\frac{2}{1+\hat\delta}{\hat\delta'^2}-\frac{3}{2}\frac{\Omega_m}{a^2}\hat\delta^2=0,
\label{growth_a2_classI}
\ee
where 
\be
	\Omega_m(a,r)=\frac{\Omega_m^0(r)}{\Omega_m^0(r)+\Omega_\Lambda^0(r)(a/a_0)^3+\Omega_k^0(r)(a/a_0)},
\label{OmegaM0_classI}
\ee
\be
	\Omega_\Lambda(a,r)=\frac{\Omega_\Lambda^0(r)(a/a_0)^3}{\Omega_m^0(r)+\Omega_\Lambda^0(r)(a/a_0)^3+\Omega_k^0(r)(a/a_0)},
\label{OmegaV0_classI}
\ee
$\Omega_k^0(r)=1-\Omega_m^0(r)-\Omega_\Lambda^0(r)$, and $a_0=a(t_0,r)$. The superscript naught for cosmological parameters denotes present day values. Eqs. (\ref{OmegaM0_classI}) and (\ref{OmegaV0_classI}) are then to be evaluated at $r_s$ and substituted into Eq.(\ref{growth_a2_classI}). We recall that the Szekeres models have 6 arbitrary functions that represent 5 physical degrees of freedom, plus a coordinate freedom to rescale $r$, and we use them as follows: 3 degrees of freedom are in $\Omega_m$, $\Omega_{\Lambda}$, and the implicit choice of a uniform bang time, whereby the scale function $a$ is determined; 2 other degrees are in setting the initial conditions of $\hat{\delta}$ and $\hat{\delta}'$ evaluated at $a$ close to zero. The initial conditions are chosen so that the growth rate starts at the usual Einstein--de Sitter matter dominated limit. The $r$-coordinate freedom is fixed by our normalization of $a$ in that for the $r_s$ used, we have set $a(t_0,r_s)=1$.

Our results for numerical integrations of the growth rate of the Szekeres with and without a $\Lambda$, as well as the $\Lambda$CDM model, are plotted in Fig.\ref{fig:luminosity}(b). The Szekeres models without a $\Lambda$ do not show any late-time suppression of the growth. We explored the parameter space of $\Omega_m$ (thus also $\Omega_k$) from 0.01 to 1.00 and found that no late-time suppression is produced. One can also note that the strong growth in such Szekeres with no $\Lambda$ should give rise to measurable effects on CMB anisotropies. 

Next, we compare the growth factor, $f=d \ln\delta/d\ln a$, in these models to current data from redshift space distortions, e.g. \cite{Blake,Guzzo2008}. We use our previous framework developed in \cite{IP2012, PIT2012} and late-time growth data bins ($a \ge 0.6$, i.e. the transition redshift). Our results are shown in Fig.\ref{fig:luminosity}(c). We find that the Szekeres model with $\Lambda$ and the $\Lambda$CDM are consistent with the growth data at $99.96\%$ and $99.99\%$ level of confidence, respectively, while the Szekeres model without $\Lambda$ provides a very poor fit to the growth data with only $31.31\%$ consistency level. Furthermore, we find that a simultaneous fit to the supernova and growth data gives $99.65\%$ level of consistency for the $\Lambda$CDM and $60.46\%$ for the Szekeres without $\Lambda$. 

It is important to provide a physical explanation to this incompatibility {between} the Szekeres models without $\Lambda$ {and} the growth data and its late-time suppression. In fact, the dynamics of a cosmological model can be fully described in terms of a set of evolution equations \cite{Ellis2}, and one of them is the Raychaudhuri equation provided earlier as Eq.(\ref{Ray2}). This equation is considered the basic equation of gravitational attraction and clustering, e.g. \cite{Ellis2,Raychaudhuri}. Now, it can be seen from this equation that the shear and gravitational tidal field that are present in general inhomogeneous cosmological models act as an effective source that adds to the matter density, and thus enhances the gravitational attraction and the growth rather than suppressing them. This is in contrast to the $\Lambda$-term that enters this equation with the opposite sign and provides a repulsive effect. 

Finally, it is worth discussing possible effects from model assumptions made by other works for the growth data reduction. Mainly, two methods have been employed to measure the growth factor. The first is a direct measurement of peculiar velocities, which are related to the underlying galaxy density measured from a redshift survey. The relation used between the peculiar velocities and the underlying galaxy density requires an angular diameter-redshift relationship, which is dependent on the $\Lambda$CDM \cite{Guzzo2008}. The second method is based on fitting the observed galaxy spectrum where the angle-redshift survey cone is mapped into a cuboid of comoving coordinates using distances from a fiducial $\Lambda$CDM model. The assumption of a $\Lambda$CDM model goes into these two methods via calculations of the angular distances as function of the redshift.  But from the first part of our analysis, the Szekeres model found has distances as function of the redshift that are practically indistinguishable from the $\Lambda$CDM one. The same holds for volume averages. It is therefore expected that the assumed $\Lambda$CDM distances in the growth data reduction would not alter the basic finding of absence of growth suppression in the Szekeres models without $\Lambda$. It is prudent though to mention that it is possible that other unforeseen systematics or assumptions have not been considered here.
%

{\parindent0pt\it \textbf{Conclusion.}}
%
We find that an analysis of the growth rate of large-scale structure in non-spherical and non-axial inhomogeneous models like the Szekeres models uncovers a serious challenge to the question of apparent acceleration associated with large-scale inhomogeneities in the universe. Szekeres models can fit well the expansion history, which we have demonstrated with an example model. However, they are found to fail to produce the observed late-time suppression of the growth unless a cosmological constant is added to the field equations. This is shown from their inconsistency with the growth data and also explained by a physical argument.   

%
{\parindent0pt\it \textbf{Acknowledgment.}}
MI acknowledges that this material is based upon work supported in part by NASA under grant NNX09AJ55G.
%

%
\end{document}